\documentclass[prl,floatfix,reprint,twocolumn]{revtex4}
\usepackage{graphicx,amssymb,amsmath,url}

\begin{document}

\title{Scaling:\\ Lost in the smog}

\author{R\'emi Louf}
\email{remi.louf@cea.fr}
\affiliation{Institut de Physique Th\'{e}orique, CEA, CNRS-URA 2306, F-91191, 
Gif-sur-Yvette, France}

\author{Marc Barthelemy}
\email{marc.barthelemy@cea.fr}
\affiliation{Institut de Physique Th\'{e}orique, CEA, CNRS-URA 2306, F-91191, Gif-sur-Yvette, France}
\affiliation{Centre d'Analyse et de Math\'ematique Sociales, EHESS-CNRS (UMR 8557),190-198 avenue de France, FR-75013 Paris, France}

\begin{abstract}
\end{abstract}

\maketitle

The success of natural sciences lies in their great emphasis on the role of quantifiable data and their interplay with models. Data and models are both necessary for the progress of our understanding: data generate stylized facts and put constraints on models. Models on the other hand are essential to comprehend the processes at play and how the system works. If either is missing, our understanding and explanation of a phenomenon are questionable. This
issue is very general, and affects all scientific domains, including the study of cities. \\

Until recently, the field of urban economics essentially consisted in untested laws and theories, unjustified concepts that supersede empirical evidence~\cite{Bouchaud:2008}. Without empirical validation, it is not clear what these models teach us about cities. The tide has turned in recent years, however: the availability of data is increasing in size and specificity, which has led to the discovery of new stylized facts and opened the door to a new science of cities~\cite{Batty:2013}. The recent craze for scaling laws~\cite{Batty:2008,Bettencourt:2007,Pumain:2004}, for instance, has been an important new step in the study of urban systems.

These laws present themselves as a power-law relationship between socioeconomic (GDP, number of patents), structural (length of roads, of cables) quantities $Y$, and the size of the population $P$ of the city:

\begin{equation}
Y = P^{\, \beta}
\end{equation}

where the exponent $\beta$ can be different from $1$. This type of scaling relation is a signature
of various processes governing the phenomenon under study, especially when the exponent
$\beta$ is not what is naively expected~\cite{Barenblatt:2003}. However, as more and more scaling
relationships are being reported in the literature, it becomes less and less clear what we really
learn from these empirical findings. Mechanistic insights about these scalings are usually
nonexistent, often leading to misguided interpretations.\\

\begin{figure}[!h]
	\centering
	\includegraphics[width=0.49\textwidth]{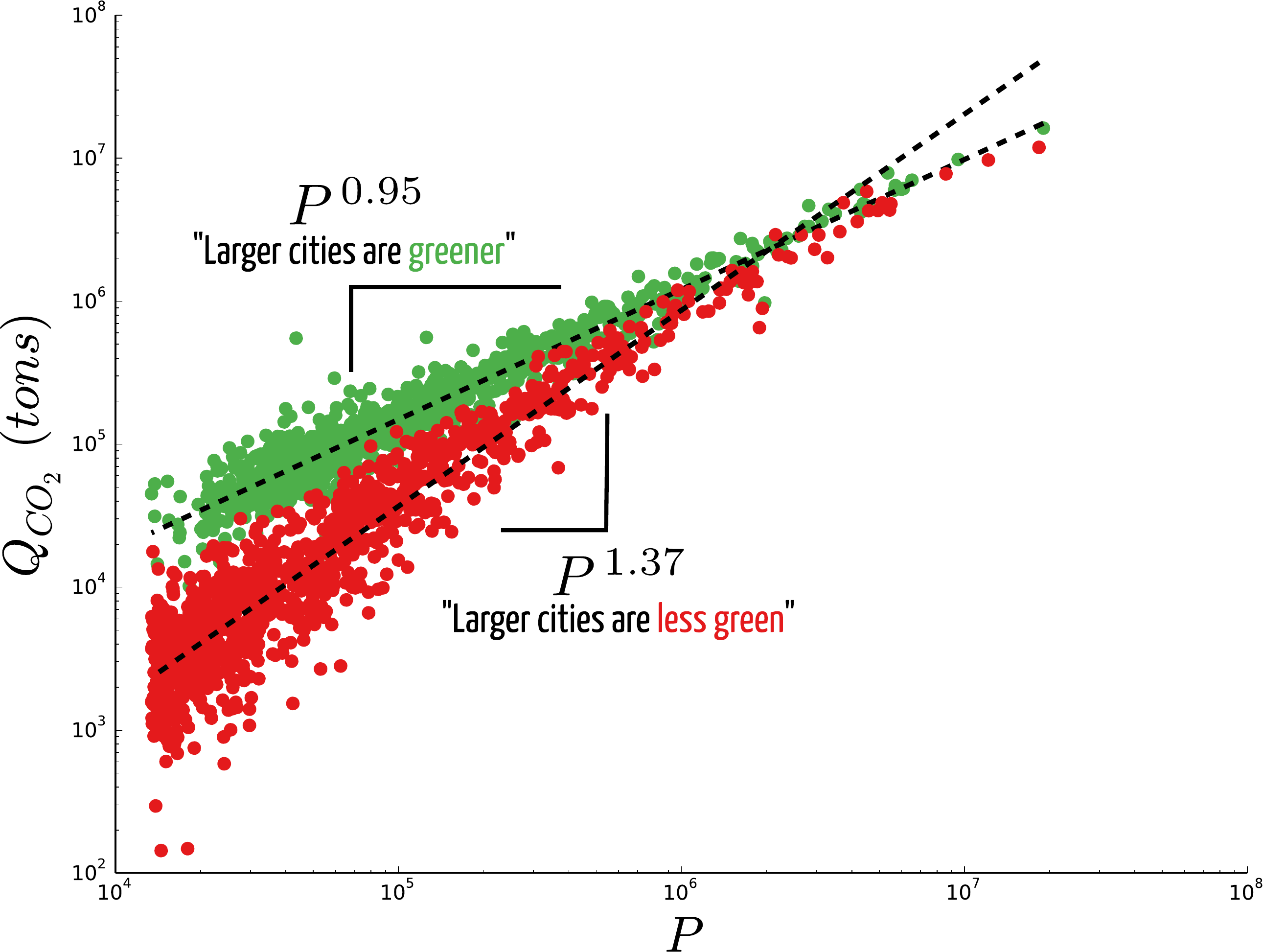}
	\caption{ {\bf Are larger cities greener or smoggier?} Scaling of transport-related $CO_2$
emissions with the population size for US cities from the same dataset but at different aggregation
levels. In red, the aggregation is done at the level of urban areas and in green for combined statistical
areas. Depending on the definition of the city, the scaling exponents are qualitatively different, leading
to two opposite conclusions. Data on $CO_2$ emissions were obtained from the Vulcan Project (\url{http://
vulcan.project.asu.ed}) (see~\cite{Fragkias:2013,Oliveira:2014}). Data on the population of urban areas and
metropolitan statistical areas were obtained from the Census Bureau (\protect\url{http://www.census.org}). \label{fig}}
\end{figure}

A striking example of the fallacies which hinder the interpretation and application
of scaling is given by different studies on $CO_2$ emissions due to transportation~\cite{Fragkias:2013,Glaeser:2010,Oliveira:2014,Rybski:2013}. The topic
is particularly timely: pollution peaks occur in large cities worldwide with a seemingly
increasing frequency, and are suspected to be the source of serious health problems~\cite{Bernstein:2004}. Glaeser and Kahn~\cite{Glaeser:2010}, Rybski et al~\cite{Rybski:2013}, Fragkias et al~\cite{Fragkias:2013}, and Oliveira et al~\cite{Oliveira:2014} are interested in how $CO_2$ emissions scale with the population size
of cities. The question they ask is simple: Are larger cities greener---in the sense that there
are fewer emissions per capita for larger cities---or smoggier? Surprisingly, these different
studies reach contradictory conclusions. We identify here two main sources of error which
originate in the lack of understanding of the mechanisms governing the phenomenon.

The first error concerns the estimation of the quantity $Q_{CO_2}$ of $CO_2$ emissions due to
transportation. In the absence of direct measures, Glaeser and Kahn~\cite{Glaeser:2010} have chosen
to use estimations of $Q_{CO_2}$ based on the total distance traveled by commuters. This is in fact
incorrect, and in heavily congested urban areas the relevant quantity is the total time spent
in traffic~\cite{Louf:2013}. Using distance leads to a serious underestimation of
$CO_2$ emissions: the effects of congestion are indeed strongly nonlinear, and the time spent
in traffic jams is not proportional to the traveled distance. As a matter of fact, commuting
distance and time scale differently with population size, and the time spent commuting and
$CO_2$ emissions scale with the same exponent~\cite{Louf:2014}.

The second, subtler, issue lies in the definition of the city itself, and over which
geographical area the quantities $Q_{CO_2}$ and $P$ should be aggregated. There is currently great
confusion in the literature about how cities should be defined, and scientists, let alone the
various statistical agencies in the world, have not yet reached a consensus. This is a crucial
issue as scaling exponents are very sensitive to the definition of the city~\cite{Arcaute:2013}. $CO_2$ emissions are no exception: aggregating over urban areas or metropolitan
statistical areas entails radically different behaviours (see figure~\ref{fig}). For the US, using the
definition of urban areas provided by the Census Bureau (\url{http://www.census.org}), one finds
that $CO_2$ emissions per capita sharply increase with population size, implying that larger
cities are less green. Using the definition of metropolitan statistical areas, also provided by
the Census Bureau, one finds that $CO_2$ emissions per capita decrease slightly with population
size, implying that larger cities are greener.\\

Faced with these two opposite results, what should one conclude? Our point is that, in
the absence of a convincing model that accounts for these differences and how they arise,
nothing. Scaling relationships, and more generally data analysis, have an important role
to play in the rising new science of cities. But, as the previous discussion illustrates, it is
dangerous to interpret empirical results without any mechanistic insight. Conclusions cannot
safely be drawn from data analysis alone.

From a policy point of view, now, what should one do to curb CO2 emissions? Favour
the growth of large urban areas or the repartition of population in less populated cities?
Both can be argued by considering data analysis alone. It should therefore be obvious that,
until they have a satisfactory understanding of the mechanisms responsible for the observed
behaviours, scientists should refrain from giving policy advice that might have unforeseen,
disastrous consequences. If they choose to do so anyway, policy makers should be wary
about what is, at best, a shot in the dark

\end{document}